  \documentclass{emulateapj}   \usepackage{apjfonts}
\usepackage{amsmath}
\newcommand{\spitzer}{\textit{Spitzer}}

\newcommand{\chandra}{\textit{Chandra}}

\newcommand{\lsim}{\lesssim}
\newcommand{\gsim}{\gtrsim}

\newcommand{\eg}{e.g.}

\newcommand{\msol}{\hbox{$\mathcal{M}_\odot$}}

\newcommand{\infinity}{\hbox{$\infty$}}

\newcommand{\uJy}{\hbox{$\mu$Jy}}
\newcommand{\ujy}{\hbox{$\mu$Jy}}

\newcommand{\mone}{\hbox{$[3.6]$}}
\newcommand{\mtwo}{\hbox{$[4.5]$}}
\newcommand{\mthree}{\hbox{$[5.8]$}}
\newcommand{\mfour}{\hbox{$[8.0]$}}

\shorttitle{CLUSTERING OF DISTANT GALAXY CLUSTERS}
\shortauthors{PAPOVICH}

\begin{document}

\slugcomment{Accepted for Publication in the Astrophysical Journal}
\title{THE ANGULAR CLUSTERING OF DISTANT GALAXY CLUSTERS\altaffilmark{1}}


\author{\sc Casey Papovich\altaffilmark{2}}
\affil{Steward Observatory, University of Arizona, 933 N.~Cherry Ave., Tucson, AZ 85721; papovich@as.arizona.edu}
\altaffiltext{1}{This work is based in part on observations
made with the \textit{Spitzer Space Telescope}, which is operated by
the Jet Propulsion laboratory, California Institute of Technology,
under NASA contract 1407}
\altaffiltext{2}{Spitzer Fellow; also Department of Physics, Texas
A\&M University, College Station, TX 77843-4242, papovich@physics.tamu.edu}


\setcounter{footnote}{2}

\begin{abstract}
We discuss the angular clustering of galaxy clusters at $z > 1$
selected within 50~deg$^2$ from the \spitzer\ Wide-Infrared
Extragalactic survey.  We employ a simple color selection to identify
high redshift galaxies with no dependence on galaxy rest--frame
optical color using \spitzer\ IRAC 3.6 and 4.5~\micron\ photometry.
The majority ($>$90\%) of galaxies with $z > 1.3$ are identified with
$(\mone -\mtwo)_\mathrm{AB} > -0.1$~mag.    We identify candidate
galaxy clusters at $z > 1$ by selecting overdensities of $\geq$26--28
objects with $\mone - \mtwo > -0.1$~mag within radii of
1.4~arcminutes, which corresponds to $r < 0.5\; h^{-1}$~Mpc at
$z=1.5$.   These candidate galaxy clusters show strong angular
clustering,  with an angular correlation function represented by
$w(\theta) = (3.1\pm0.5) (\theta/1\arcmin)^{-1.1\pm0.1}$  over scales
of 2--100~arcminutes.   Assuming the redshift distribution of these
galaxy clusters follows a fiducial model, these galaxy clusters have
a spatial--clustering scale length $r_0 = 22.4\pm 3.6\, h^{-1}$~Mpc,
and a comoving number density $n=1.2\pm 0.1 \times 10^{-6} h^{3}$~Mpc$^{-3}$.
The correlation scale length and number density of these objects are
comparable to those of rich galaxy clusters at low redshift.   The
number density of these high--redshift clusters correspond to
dark--matter halos larger than $3-5\times 10^{13}$ $h^{-1}$ \msol\ at $z=1.5$.
Assuming the dark halos hosting these high--redshift clusters grow
following $\Lambda$CDM models, these clusters will reside in halos
larger than
$1-2\times 10^{14}$ $h^{-1}$ \msol\ at $z=0.2$, comparable to rich
galaxy clusters.
\end{abstract}
 
 

\section{Introduction}\label{section:intro}

Clusters of galaxies provide important samples for studying
structure evolution and cosmology.  They trace large dark mass halos,
which collapsed early in the history of the Universe, and thus they probe
the structure of overdensities in the underlying dark--matter
distribution \citep{spr05}.   This is evident in the strong spatial
clustering inferred for galaxy cluster samples, which find typical spatial
correlation function scale lengths of $r_0 \sim 20-30$~$h^{-1}$ Mpc
for optically and X-ray selected galaxy clusters in the local and
distant Universe
\citep[e.g., ][]{bah88, aba98, lee99,col00,gon02,bah03,bro07}.    Furthermore,
the expected number density of galaxy clusters is sensitive to the
cosmic matter density, $\Omega_m$ \citep[\eg,][]{kit96,wan98}.  Currently, both the
measured correlation function scale lengths and number densities of galaxy
clusters support theoretical predictions for  standard cold
dark--matter models, which include a cosmological constant
\citep[\eg,][]{bah03}.  Therefore, because galaxy clusters correspond to
large matter overdensities, their number density evolution with
redshift provides constraints on cosmological
parameters, and these constraints should be amplified at higher
redshifts \citep[e.g., when the Universe was matter dominated; see ][]{hai01}.

Galaxy clusters also provide laboratories for studying galaxy
formation.  Galaxy clusters contain a population of
galaxies that evolved early in the history of the Universe.   Locally,
clusters contain a high fraction of early--type, elliptical and
lenticular galaxies, which contain little ongoing star formation
\citep[\eg,][]{dre80}, and the fraction of early--type galaxies in
clusters appears to evolve strongly with redshift
\citep[\eg,][]{lub98,vandok00}.   Studies of the stellar populations
of the early--type cluster galaxies at $z \lsim 1$ show they have evolved
nearly passively from $z_f \sim 2-3$
\citep[\eg,][]{pos98,sta98,vandok01,vandok07}.   Studying clusters at
$z\gsim 1$ provides constraints on the formation of the massive,
early--type galaxies within them \citep[\eg,][]{eis07}.

Identifying galaxy clusters at $z\gsim 1$ presents some significant
technical and physical challenges.  Deep X-ray surveys identify
distant clusters at cosmologically significant redshifts, because the
X-ray luminosity scales with the mass of the cluster and it is
relatively unaffected by projection effects \citep[\eg,][]{ros04}.
Spectroscopic observations of galaxies associated with faint, diffuse
X--ray emission have identified clusters at redshifts (to date) of
$z = 1.41$--1.45 \citep{mul05,sta06}.  However, the X-ray surface brightness
declines strongly with redshift ($\propto [1+z]^4$), biasing against
high--redshift clusters with relatively diffuse emission.
Furthermore,  X--ray emission  from hot ($T\sim 10^{6}-10^{7}$~K) gas
in the inter-cluster medium generally requires a dynamically relaxed,
virialized, massive system.  This may not be the case at high
redshifts during the hierarchical assembly of the dark matter halos
where cluster progenitors will be less massive and likely disrupted
\citep[\eg,][]{ros02}.    Searches for high--redshift clusters have also
targeted fields around distant radio galaxies as ``signposts'' of
large dark--matter overdensities
\citep{kur00,ven02,ven07,ste03,mil04,cro05,kaj06,kod07}.  While these have
been successful, these studies require the presence of a massive,
central galaxy,  which may not be an intrinsic, ubiquitous feature to
the high--redshift progenitors of all clusters.

Using ground--based optical and near--IR, or \spitzer\ near--IR imaging, recent
searches for clusters rely on identifying overdensities of galaxies
with red optical to near--IR colors
\citep[\eg,][]{gla05,gla07,kaj06,wilson06,kod07}, which correspond to galaxies
with evolved stellar populations with strong Balmer/4000~\AA\ breaks
at $z\gsim 1$.  While these surveys have had success, the cluster
selection  based on red galaxy colors may be biased away from
potential clusters dominated by galaxies with relatively blue colors.
For example, any study of the evolution of galaxies in these clusters
are subject to a form of ``progenitor'' bias in that one identifies
only those clusters dominated by red, presumably early-type galaxies
\citep[see, \eg,][]{vandok01}.   Because these red, early-type
galaxies that dominate galaxy clusters  at present formed their
stellar populations at $z\gsim 1$ (see references above), one expects
that their progenitors should show increasing indications of ongoing
star formation at these epochs.   Furthermore, there is 
evidence that the relative fraction of galaxies with blue colors in overdense
environments increases with redshift \citep{ger07}, with luminous blue
galaxies at $z\sim 1$ preferentially residing in regions of
greater--than--average overdensity \citep{coo07}.   For example,
\citet{elb07} found that the average star formation rates of galaxies
at $z\sim 1$ in dense environments are higher than those of other
co--eval galaxies in less dense environments.   Certainly by $z\gsim
1.5$ there exist strong overdensities of blue, rest-frame UV--selected
galaxies \citep{ste98,ade05b,steidel05}, and such systems would
presumably be missed by traditional searches for red galaxies.  For
example, \citet[see also Eisenhardt et al.\ 2007]{bro06} and
\citet{vanbreukelen06} identify
high--redshift cluster candidates as overdensities of galaxies with
similar photometric redshifts, and these efforts have yielded some of the
highest redshift clusters yet identified at $z=1.41$ \citep{sta05} and
$z=1.45$ \citep{vanbreukelen07}.

Here, we use a simple color selection to identify galaxies at $z\gsim
1$ based solely on red colors between IRAC channels 1 and 2 (3.6 and
4.5~\micron).  This selection has little dependence on galaxy
rest-frame optical color, it is sensitive to red and blue galaxies
nearly equally.  This technique utilizes the fact that nearly all
plausible stellar populations show a peak in their $f_\nu$ emission at
1.6~\micron\, accompanied by a steady decline on the Rayleigh--Jeans
tail of the stellar emission (see Simpson \& Eisenhardt 1999; Sawicki
2002; van Dokkum et al.\ 2007).     All composite stellar populations
in galaxies at $z\lsim 1$ have blue IRAC $\mone - \mtwo$ colors
because these bands sit on the Rayleigh--Jeans tail of the stellar
emission, while at $z\gsim 1$ galaxies appear redder in IRAC $\mone -
\mtwo$ color as these bands probe the peak of the stellar emission.   A
similar technique has been applied by \citet{tak07} to identify
galaxies at $z\gsim 0.5$ using $[2.2]-[3.5] > 0.1$~mag colors from the
AKARI satellite, analogous to the technique used here.  We identify
candidates for high--redshift, $z\gsim 1$, galaxy clusters from
overdensities of those galaxies with IRAC $\mone - \mtwo > -0.1$~mag
colors.   Strictly speaking, these galaxy overdensities are candidates
for galaxy clusters, and must be confirmed by spectroscopy.
Nevertheless, we show that even these candidates for galaxy clusters
provide a useful sample for studying galaxy evolution and cosmology at
high redshifts.

A strong motivation for this study is the proposed
\spitzer\ mission in the post--cryogenic era (starting ca.\ 2009 April),
which  may provide a tremendous amount of observing time ($>$$10^4$
hrs) with only IRAC channels 1 and 2.   This study demonstrates one
possible science driver for a warm \spitzer\ mission \citep[see
also][]{vandok07b}. The outline for the rest of this paper is as
follows.  In \S~2 we describe the datasets used for the study.  In
\S~3 we describe the method to identify distant galaxy clusters from
\spitzer/IRAC data.  In \S~4 we compute the angular clustering of
the distant galaxy clusters.  In \S~5 we discuss the redshift
selection function and number density of these clusters, we derive
their spatial clustering correlation length from the angular
correlation function, and we describe the evolution of these galaxy
clusters. Throughout this work we quote optical and near--IR
magnitudes on the AB system where $m_\mathrm{AB} = 23.9 - 2.5 \log(
f_\nu/1\;\uJy)$.    We denote magnitudes measured from the data with
\spitzer/IRAC in the four channels [3.6], [4.5], [5.8], [8.0],
respectively.   Throughout, we use a cosmology with $\Omega_m = 0.3$,
$\Lambda = 0.7$, and $H_0 = 100\, h$~km s$^{-1}$ Mpc$^{-1}$.   To
compare with other results, we assume a Hubble parameter $h=1.0$.

\begin{figure*}[t]
\plotone{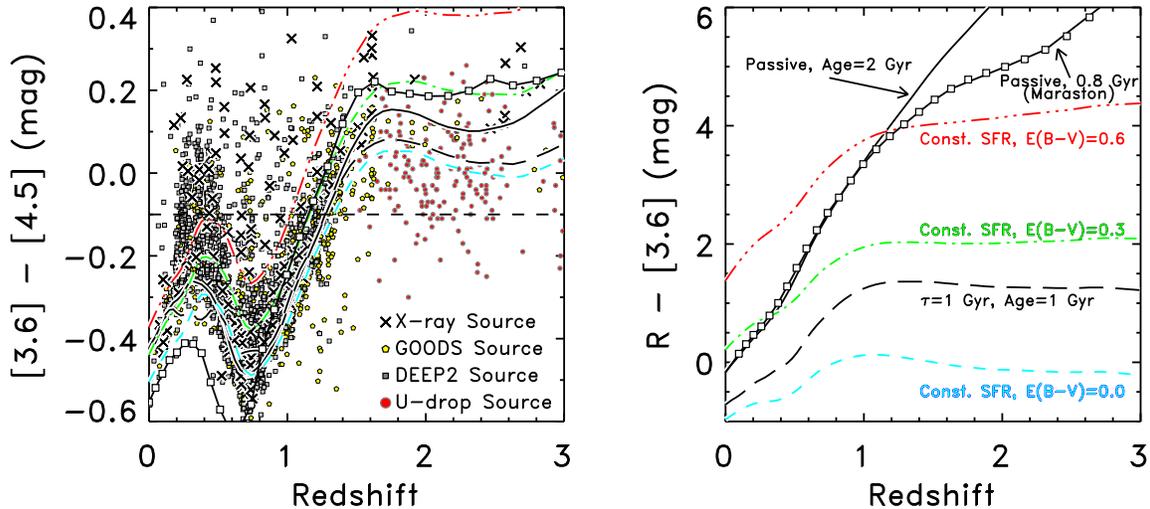}
\caption{ Galaxy colors as a function of redshift.  The left panel
shows the $\mone - \mtwo$ color as a function of redshift, and the
right panel shows the $R-\mone$ color as function of redshift.   The
data points for $\sim$10$^4$ objects are shown only in the left panel for clarity.  The 
pentagons and squares denote sources from GOODS and DEEP2 with
high--quality spectroscopic redshifts.  Crosses show those sources
detected in X-ray data.    Red circles show blue, star--forming
galaxies with spectroscopic redshifts $1.5 < z < 3$ from
\citet{sha05} and \citet{red06}.  The horizontal dashed line shows the $\mone - \mtwo >
-0.1$~mag color criterion.    The curves show the colors expected from
various integrated stellar population models, labeled in the right
panel.  The white boxes connected by the solid line shows the colors expected for a 
passively evolving stellar population with $t=0.8$~Gyr from
\citet{mara06}.  The other curves assume the \citet{bru03} model.  The solid
line corresponds to a passively evolving stellar population with an
age $=2$~Gyr.  The long--dashed line corresponds to a stellar
population with an exponentially declining star--formation rate,
$\exp( -t/\tau)$ with $\tau=t=1$~Gyr.  The short--dashed, dot--dashed,
and triple-dot--dashed lines correspond to a stellar population with
constant star--formation and dust extinction of $E(B-V)$=0.0, 0.3, and
0.6, respectively, assuming the \citet{cal00} extinction law. 
\label{fig:zvcolor}}
\end{figure*}

\section{The Datasets}\label{section:data}

To demonstrate the utility of using IRAC colors to identify galaxies
with $z\gsim 1$, we use two datasets with ample spectroscopic
redshifts and deep IRAC imaging.  These are the southern Great
Observatories Origins Deep Survey (GOODS--S; Dickinson et al.\ 2003;
Giavalisco et al.\ 2004) and the All-wavelength Extended Groth strip
International Survey (AEGIS; Davis et al.\ 2007).   Although other
datasets with comparable IRAC data and spectroscopic data exist, they
would add little to the discussion here.   However, we do include the
samples of UV--bright, star--forming galaxies  (so--called
``U--dropouts'') with $1.5 < z < 3$ from \citet{sha05} and
\citet{red06} to augment the galaxies with IRAC data and spectroscopy
at higher  redshifts. For the bulk of the study here, we use available
catalogs from the \spitzer\ Wide-Infrared Extragalactic (SWIRE)
survey, which cover the largest area with deep \spitzer\ data.

\subsection{The GOODS dataset}

GOODS--S includes deep \spitzer/IRAC imaging to a depth of 25~hr.\ per
band at 3.6, 4.5, 5.8, and 8.0~\micron\ over a $10\arcmin \times
15\arcmin$ in the southern Chandra Deep Field.    These data reach
limiting $5\sigma$ flux sensitivities for point sources of 0.11 and
0.21~\ujy\ at 3.6 and 4.5~\micron, respectively (M.~Dickinson et al.\
2007, in preparation).  The GOODS--S field has extensive spectroscopy.
Here we use the published redshifts of \citet{lefevre04},
\citet{szo04}, \citet{mig05}, and \citet{van05,van06}, which provide
1624 redshifts for GOODS IRAC sources.  For this field we also make
use of the X-ray catalog from deep (1~Msec) \chandra\ data (Alexander
et al.\ 2003), which provides X-ray counterparts to 133 of the IRAC
sources with spectroscopic redshifts.

\subsection{The AEGIS dataset} 

The AEGIS program includes deep \spitzer/IRAC imaging to a depth of
3~hr.\ per band at 3.6, 4.5, 5.8, and 8.0~\micron\ over a $15\arcmin
\times 2^\circ$ field in the Groth Strip (Davis et al.\ 2007).  These data reach
$5\sigma$ depths of 0.9~\ujy\ at 3.6 and 4.5~\micron\ \citep{barmby07}.  This field covers the
DEEP2 spectroscopic survey using Keck/DEIMOS, and provides redshifts
for $>$8000 IRAC sources.  Spectroscopic targets for DEEP2 extend to
sources with $R < 24.1$~mag, although the $R$--band
imaging identifies sources to a limiting magnitude of $R = 24.7$~mag
($5\sigma$; see, Davis et al.\ 2007).
We also use the X-ray catalog from 200~ksec
data in this field \citep{nan05}, which provides X-ray
counterparts to 32 IRAC sources with spectroscopic redshifts.

\subsection{The SWIRE dataset}\label{section:SWIRE}

For the work here, we use data from the SWIRE legacy survey \citep{lon03}.
SWIRE covers six fields with \spitzer/IRAC to 120~s depth, reaching
estimated $5\sigma$ flux limits of 3.7 and 5.4~\ujy\ at 3.6 and
4.5~\micron, respectively.   The SWIRE fields cover roughly 50~deg$^2$
in total, including 7.8~deg$^2$ in the southern \chandra\ Deep Field
(CDF--S), 11.0~deg$^2$ in the Lockman Hole, 9.2, 4.8, and 6.9~deg$^2$
in the N1, N2, and S1 fields of the European Large-area \textit{ISO}
Survey (ELAIS)  fields, and 9.2~deg$^2$ in the XMM Large--Scale Survey
(XMM-LSS), respectively.\footnote{see
http://swire.ipac.caltech.edu/swire/astronomers/}    For the study
here, we used the SWIRE IRAC data from the publicly available third
data release (DR3), as described in \citet{sur05}\footnote{see also
http://irsa.ipac.caltech.edu/data/SPITZER/SWIRE/}.

\section{The High Redshift Galaxy Cluster Sample}\label{section:selection}

\subsection{IRAC Color Selection of High--Redshift Galaxies}\label{section:colorselection}

Based on our understanding of the emission of composite stellar populations,
galaxies at high redshift should have relative uniformity
in their IRAC $\mone - \mtwo$ colors.   Figure~\ref{fig:zvcolor} shows
the expected behavior of the $\mone -\mtwo$ color as a function of
redshift for various stellar population models.    The curves in the
figure correspond to a wide range of composite model stellar
populations, using both the \citet{bru03} models
(including dust extinction from Calzetti et al.\ 2000), and the
\citet{mara06} models.  The diversity in the models is reflected
in the variation in their $R - \mone$ colors, shown in the
right panel of figure~\ref{fig:zvcolor}.  These  models include purely
passive, old stellar populations formed in an instantaneous burst,
models with exponentially declining star formation rates, and models
with constant star--formation rates and various dust attenuation.     Even
thought the models are diverse, they span a tight locus in $\mone -
\mtwo$ color, showing a characteristic ``S'' shape with a local
maximum at $z\sim 0.3$, a local minimum at $z\sim 0.7$, and a rise to
red colors for $z\gsim 1$.     For $z\lsim 1$ the composite stellar
population models have relatively blue $\mone - \mtwo$ colors as these
are dominated from stellar emission at $\lambda > 2$~\micron, which
corresponds to the Rayleigh--Jeans tail of stellar spectra.  The rise
at $z\gsim 1$ results as the IRAC \mone\ and \mtwo\ bands shift to
wavelengths $\lsim 2$~\micron, where they probe the peak in the
stellar emission (in $f_\nu$ units) near 1.6~\micron\ (e.g., Sawicki
2002).   At yet higher redshifts, IRAC probes $\lambda < 1$~\micron,
where composite stellar populations have relatively red $\mone -
\mtwo$ colors. 

\begin{figure}[t]
\epsscale{1.1}
\plotone{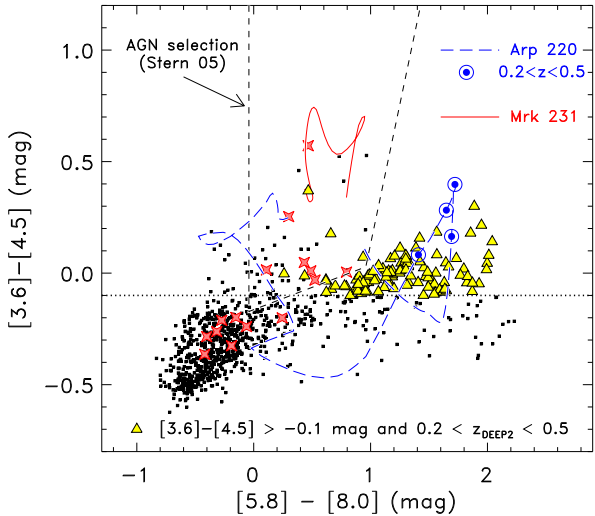}
\epsscale{1.0}
\caption{IRAC $\mthree - \mfour$ versus $\mone-\mtwo$  colors for
objects in the AEGIS field with high--quality spectroscopic redshifts
and high--S/N IRAC photometry.   Yellow triangles denote those objects
with $0.2 < z_\mathrm{DEEP2} < 0.5$ and $\mone - \mtwo > -0.1$~mag
(indicated by the dotted line).   Such objects do not generally
populate the mid--IR color selection for obscured AGN proposed by
\citet{ste05}, bounded by the dashed region.  The colored curves show
the expected colors as a function of redshift for the IR--luminous
galaxies Arp~220, dominated by star--formation (blue, dashed line),
and Mrk~231, which hosts an obscured AGN (red solid line).     Red
stars show X-ray--detected sources. Sources denoted by the yellow
triangles have red IRAC colors consistent with warm dust associated
with star-formation in objects with $0.2 < z < 0.5$ such as Arp~220
(the redshifts $0.2 < z < 0.5$ for Arp~220 are indicated by large blue
circles). \label{fig:iraccolors}}.
\end{figure}

Existing IRAC data show that high--redshift galaxies generally have
$\mone-\mtwo$ colors within expectations from the composite stellar
population models.  Figure~\ref{fig:zvcolor} shows the IRAC colors for
galaxies with high spectroscopic quality in the GOODS--S and AEGIS
fields, and the blue, star--forming galaxies at higher redshift from
\citet{sha05} and \citet{red06} (see \S~\ref{section:data}).   The bulk of the data mirror the
characteristic ``S'' shape expected from the models, although there
are noticeable outliers.  Many sources with putative AGN based on the
X-ray data have red $\mone-\mtwo$ colors for all redshifts.   These
objects presumably have a contribution to their emission at
$1-3$~\micron\ arising from dust heated by an AGN.    There is also a
subset of galaxies with $0.2 < z < 0.5$ with $\mone-\mtwo$ colors
redder than predicted by any of the composite stellar population
models.    These colors do not arise from differences in the modeling
the evolution of post--main sequence stars.  Figure~\ref{fig:zvcolor}
shows the expected colors from a stellar population with enhanced
rest--frame near-IR emission from thermally pulsating asymptotic giant
branch stars \citep{mara06}.  Although the figure shows
this model for only one possible age (0.8~Gyr, near the peak
of the contribution of post--main--sequence stars to the bolometric
emission; Maraston 2005), no other possible age and star-formation
history for this model matches the points with $\mone - \mtwo >
-0.1$~mag at $0.2 < z < 0.5$.

Interestingly, the objects with $0.2 < z < 0.5$ and $\mone - \mtwo >
-0.1$~mag do not generally have colors consistent with AGN.
Figure~\ref{fig:iraccolors} shows the IRAC $\mthree-\mfour$ versus
$\mone - \mtwo$ colors for objects in the DEEP2 field with high--S/N
IRAC photometry and high--quality spectroscopic redshifts.  The
objects at $0.2 < z < 0.5$ with $\mone - \mtwo > -0.1$~mag have
$\mthree-\mfour$ colors redder than expected for obscured AGN
\citep{ste05}.   Instead, these objects have IRAC colors consistent
with IR--luminous star-forming galaxies, such as Arp~220.   In such
objects the red $\mone-\mtwo$ colors result from warm dust heated
by star--formation at rest--frame $\lambda > 2$~\micron\
\citep{ima06}.  The red $\mthree - \mfour$ colors result
at $z > 0.2$ as the strong mid--IR emission features at $\lambda >
6$~\micron\ shift out of the \mthree\ bandpass (but remain in
the \mfour\ bandpass until $z \simeq 0.5$).    Therefore, the
population of galaxies with $0.2 < z < 0.5$ with $\mone - \mtwo >
-0.1$~mag are star-forming galaxies with a strong contribution of warm
dust emission at rest--frame $\lambda > 3$~\micron\ over what is
expected from composite stellar populations.    
 
\begin{figure}[b]  
\epsscale{1.2}
\plotone{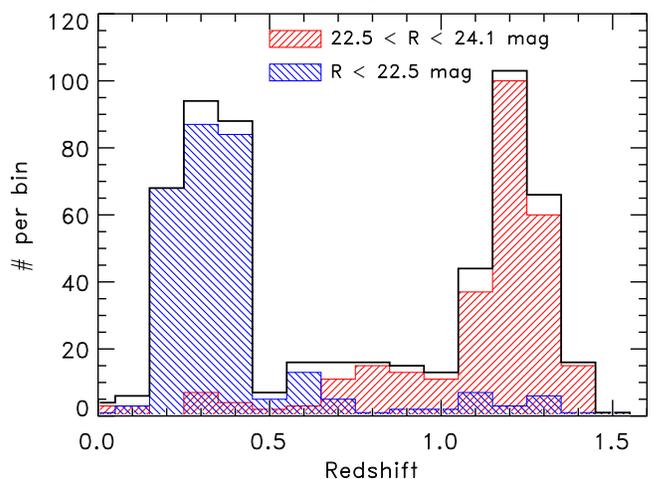}
\epsscale{1.0}
\caption{The redshift distribution of sources
with $\mone - \mtwo > -0.1$~mag for sources with high quality
redshifts from the DEEP2 spectroscopic survey.  The solid line shows
the distribution of galaxies satisfying this IRAC color cut.  The
blue--shaded and red--shaded histograms show the redshift distribution
of those galaxies with $R < 22.5$~mag and $22.5 \leq R < 24.1$~mag,
respectively.  The appreciable decline in the number of sources with
redshift at $z \gsim 1.2$ results from the spectroscopic incompleteness
of the DEEP2 survey.  \label{fig:zdist}}
\end{figure}

Based on figure~\ref{fig:zvcolor} the $\mone - \mtwo > -0.1$~mag
selection identifies galaxies with $0.2 < z < 0.5$ and $z > 1$.
Including an apparent magnitude criterion further differentiates these
galaxies.  Figure~\ref{fig:zdist} shows the redshift distribution of
the $\mone - \mtwo > -0.1$~mag galaxies with $R \leq 22.5$~mag and
$22.5 < R < 24.1$~mag  for galaxies with high--quality DEEP2
spectroscopic redshifts (where $R = 24.1$~mag is the limiting
magnitude for the DEEP2 spectroscopy).   The objects from DEEP2 in
this figure include only those with IRAC 3.6 and 4.5~\micron\ flux
densities above the SWIRE IRAC detection limit.  Thus, the redshift
distribution is applicable to the SWIRE data used below.  The vast
majority of sources with $R < 22.5$~mag also have $0.2 < z < 0.5$,
whereas most galaxies with $R > 22.5$~mag have $z > 1.1$.    The
appreciable decline in the number of sources with redshift at $z \gsim
1.2$ results from the spectroscopic incompleteness of the DEEP2
survey.   Based on figure~\ref{fig:zvcolor}, we expect that the
redshift distribution of galaxies with $\mone -\mtwo > -0.1$~mag and
$R > 22.5$~mag will certainly extend to redshifts greater than 1.4.

\begin{figure}[tb]  
\epsscale{1.2}
\plotone{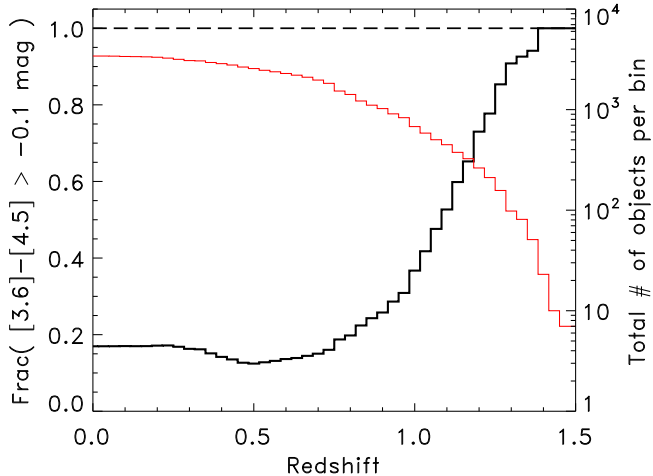}
\epsscale{1.0}
\caption{The fraction of sources with $\mone - \mtwo > -0.1$~mag as a
function of redshift from the DEEP2 spectroscopic survey.  The thin,
red line shows the total number of all objects in the DEEP2
spectroscopic catalog.  The thick, black line shows the fraction of
those sources with $\mone - \mtwo > -0.1$~mag.  More than 90\% of the
sources at $z > 1.3$ have $\mone - \mtwo >
-0.1$~mag. \label{fig:check_egs}}
\end{figure}

Even in the absence of deep optical imaging, the selection of $\mone -
\mtwo > -0.1$~mag sources preferentially identifies high--redshift
galaxies.   Galaxies with $\mone - \mtwo > -0.1$~mag and $R <
22.5$~mag constitute only $\simeq$20\%  of all objects with DEEP2
spectroscopic redshifts and IRAC 3.6 and 4.5~\micron\ flux densities
brighter than the SWIRE flux limit.    Furthermore, among all galaxies in the
AEGIS field with $\mone - \mtwo > -0.1$~mag (with or without
spectroscopic redshifts) 80\% have $R > 22.5$~mag, implying the
majority of these have $z\gsim 1$.
Therefore, galaxies with $\mone - \mtwo > -0.1$~mag account for the
majority of galaxies with $z \gsim 1$.  Figure~\ref{fig:check_egs}
shows the fraction of galaxies with $\mone - \mtwo  > -0.1$ as a
function of redshift from the high--quality spectroscopic redshifts in
the AEGIS field.  More than 50\% (90\%) of the galaxies with $z > 1.1$
(1.3) satisfy the IRAC color--selection criteria, $\mone - \mtwo >
-0.1$~mag, and this increases to 100\% for AEGIS galaxies with $z >
1.4$.  Moreover, $\simeq 90$\% of the galaxies at
$1.5 < z < 3$ from \citet{sha05} and
\citet{red06} satisfy $\mone - \mtwo > -0.1$~mag (the ``U--dropouts'',
see figure~\ref{fig:zvcolor}).  The IRAC color selection is highly
efficient at identifying high redshift galaxies.

\subsection{Identification of High Redshift Clusters}\label{section:id}

We identify candidate galaxy clusters as overdensities of galaxies
satisfying the color selection $\mone - \mtwo > -0.1$~mag in the SWIRE
data.  We take all objects in the SWIRE IRAC catalogs for each of the
six fields (see \S~\ref{section:SWIRE}) with $\mone - \mtwo >-
0.1$~mag, and  S/N$(3.6\micron) > 10$ and S/N$(4.5\micron) > 10$. 
The S/N limit ensures that only well-detected objects enter the
sample.    In practice, this limits the analysis to objects with
$f_\nu(3.6) > 7-10$~\uJy.   Using a pure flux-density limit does not
affect the results here, but increases the possibility that the sample will
contain spurious sources with low significance or uncertain IRAC
colors.    The goal here is to identify a robust set of candidate
galaxy clusters.

To define candidate galaxy clusters, we count the number of galaxies
within a given radius on the sky.   For each galaxy in the sample with
$\mone - \mtwo > -0.1$~mag, we count the number of other galaxies
satisfying this color criterion within a radius of $1\farcm4$.   For
$1 < z < 2.5$ (the approximate range the expected redshift
distribution given our selection, see \S~\ref{section:rsf}) $1\farcm4$
corresponds to an angular diameter distance of
$\approx$0.5~$h^{-1}$Mpc (for $\Omega_m = 0.3$ and
$\Lambda=0.7$). This size is smaller than the typical size used to
identify galaxy clusters in optical data (e.g., Bahcall et al.\ 2003),
but this size encompasses typical cluster--core sizes.
Moreover, the smaller radius used here reduces greatly the number of
chance alignments along the line of sight, which scale as $\propto
r^2$.   Experiments using larger radii for this selection require a
substantially greater number of sources to exceed the $3\sigma$
threshold (see below), and identify objects with greater angular clustering.

\begin{figure}[t]
\epsscale{1.2}
\plotone{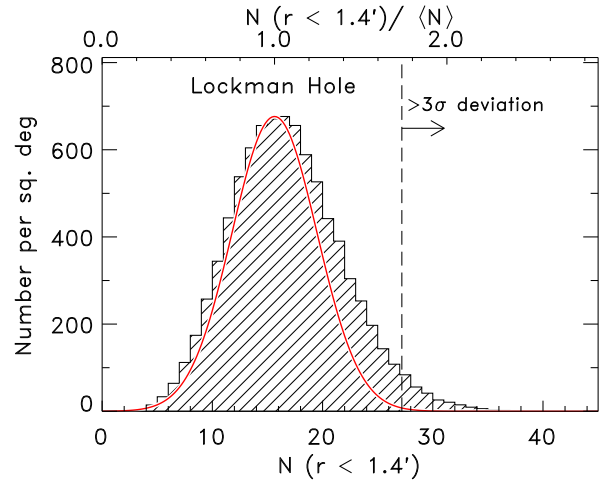}
\epsscale{1.00}
\caption{The distribution of the number of objects with $\mone - \mtwo > -0.1$~mag within
$r < 1\farcm4$ for sources in the Lockman Hole field.
The mean number of objects is $\langle N(r < 1\farcm4) \rangle =
15.6$.  The measured distribution is skewed to objects with a high number of
companions.  The red curve shows a Gaussian fit to the clipped
distribution.  The vertical dashed line shows three standard
deviations of the Gaussian distribution, corresponding to objects with $\geq 27$
companions with $\mone - \mtwo > -0.1$~mag within  $r < 1\farcm4$.  \label{fig:ndist}}
\end{figure}

Figure~\ref{fig:ndist} shows the distribution of the number of
companions with $\mone - \mtwo > -0.1$~mag within $r < 1\farcm4$ for
the Lockman Hole SWIRE field.  In this case, the mean
number of galaxies with these IRAC colors within $r <
1\farcm4$ is $\langle N \rangle \simeq 15.6$.  The distribution is
skewed toward objects with an excess number of companions (this is the
case for all the other SWIRE fields as well), which indicates the
strong clustering of these sources.    If the overdensities of objects
result from projection effects of unassociated objects along the line
of sight, or from other random processes, then the distribution in
figure~\ref{fig:ndist} would be more consistent with a Gaussian
distribution.

We fit a Gaussian to the distribution of the number of objects with
$\mone - \mtwo > -0.1$~mag within $r < 1\farcm4$ for each of the SWIRE
fields.  For each field, we fit a Gaussian to the distribution
iteratively clipping at $2\sigma$.  This fit is drawn on the
distribution in figure~\ref{fig:ndist} for the Lockman Hole, and it
matches the left--hand side of the observed distribution very well.
We then defined the sample of galaxy cluster candidates to be those
objects with more than $\langle N \rangle + 3\sigma_N$ other objects
with these IRAC colors within these radii, where $\sigma_N$ is the
width of the fitted Gaussian.   For the six fields, this corresponds
to objects with more than 26--28 companions, see table~\ref{table}.

Many of the objects with high overdensities of companions are counted
as members in multiple cluster candidates.   To remedy this, we merged
the cluster candidates by applying a friend--of--friend algorithm with
a linking-length of $1\farcm4$ to all the galaxies counted as cluster
members.   In this way, objects previously counted in more than one
candidate cluster are subsequently assigned to one and only one
cluster.  Subsequently, any clusters with less than the requisite
number of galaxies is then merged with the nearest neighoring
cluster.  In this way, all clusters have more than 27--29 objects and
each object belongs to only one cluster.

Table~\ref{table} gives the surface density of galaxy cluster
candidates, some statistics, and the areal coverage for each of the
six SWIRE fields.   For the remainder of this paper, we call these
``galaxy clusters'' or ``high--redshift galaxy clusters'' for brevity.
They are, strictly speaking, unconfirmed cluster candidates at high
redshifts ($z\gsim 1$), and require verification either by very
deep spectroscopy or accurate photometric redshifts.

\section{Angular Clustering of High Redshift Clusters}\label{section:wtheta}

\begin{figure*}[t]  
\epsscale{1.15}
\plotone{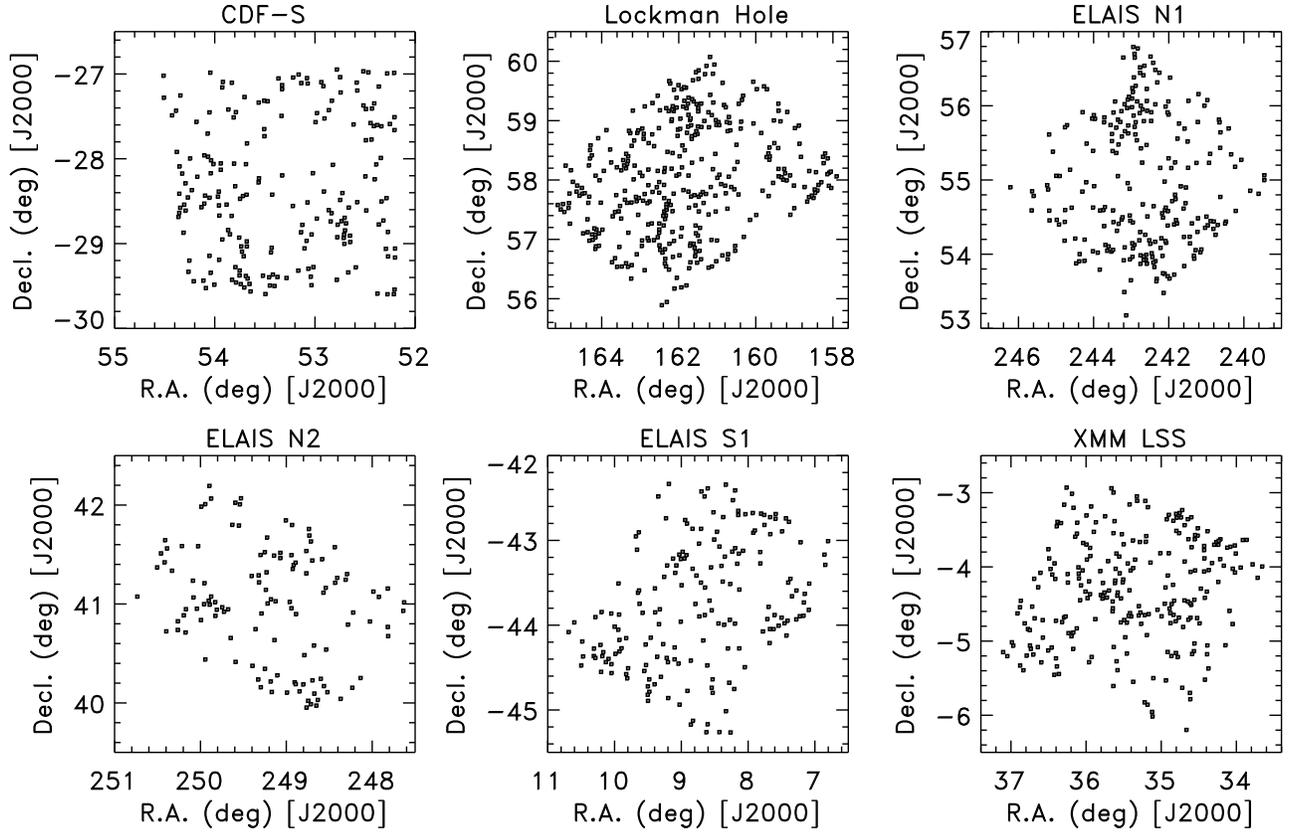}
\epsscale{1.00}
\caption{The spatial distribution of galaxy cluster candidates in the
six SWIRE fields (as labeled).  Each data point corresponds to one
cluster candidate, defined in \S~\ref{section:id}.   Each
field has very different areal coverage, see table~\ref{table}.   The
scale of the right ascension and declination varies in each
panel. \label{fig:radec}} \end{figure*}

Galaxy clusters trace massive overdensities in the dark matter
distribution, and thus galaxy clusters should show  strong spatial
clustering.   We expect that the galaxy clusters  defined in
\S~\ref{section:id}  have redshifts $z \gsim 1$ based on their $\mone -
\mtwo > -0.1$~mag color selection, and we use this galaxy cluster
sample to study the clustering of these objects at high redshift.

\begin{deluxetable}{lcccc}
\tablewidth{0pt}
\tablecaption{Statistics on Clusters in the SWIRE fields\label{table}}
\tablehead{ \colhead{} & \colhead{Area} & \colhead{} & \colhead{} & \colhead{$\mathcal{N}$} \\ 
\colhead{Field} & \colhead{(deg$^2$)} & \colhead{$\langle N(r <
1\farcm4) \rangle$} & 
\colhead{$\sigma_N$} & \colhead{(\# per deg$^2$)} \\
\colhead{(1)} & \colhead{(2)} & \colhead{(3)} & \colhead{(4)} &
\colhead{(5) } }
\startdata
CDF--S &  \phn7.8  & 15.4 & 3.9 & 25.2 \\
ELAIS N1 & \phn9.3 & 15.3 & 3.8 & 25.9 \\
ELAIS N2  & \phn4.2 & 14.9 & 3.6 & 28.6 \\
ELAIS S1  &  \phn6.8  & 15.5 & 3.8 & 26.7 \\
Lockman Hole & 11.1 & 15.6 & 3.9 & 33.2 \\ 
XMM--LSS & \phn9.1 & 16.0 & 4.0 & \phn27.3
\enddata
\tablecomments{ (1) SWIRE field name, (2) IRAC data areal coverage,
(3) mean number of companions with $\mone - \mtwo > -0.1$~mag within
$r < 1\farcm4$, (4) standard deviation of clipped Gaussian of number
of companions with $\mone - \mtwo > -0.1$~mag within $r < 1\farcm4$,
(5) Surface density of galaxy clusters. }
\end{deluxetable}

Figure~\ref{fig:radec} shows the angular distribution of the galaxy
cluster samples for each of the six SWIRE fields.   The angular
distribution of the high redshift clusters is clearly clustered, with
obvious overdensities and voids.   Even though the scale of each panel
in figure~\ref{fig:radec} varies somewhat, it is clear that the number
density of high redshift galaxy clusters shows strong field--to--field
variance, even in fields as large as those available from SWIRE
($>$5~deg$^2$).    Table~\ref{table} includes the surface densities of
high--redshift galaxy clusters in the SWIRE fields.   These vary from
$\mathcal{N} = $25.2~deg$^{-2}$ for the CDF--S field to
33.2~deg$^{-2}$ for the Lockman Hole field.   The mean surface density for
all fields combined is $\langle \mathcal{N}
\rangle$=28.1~deg$^{-2}$ with a standard deviation of 0.3~deg$^{-2}$ from
counting statistics only.   This uncertainty is significantly smaller than
the field--to--field standard deviation, $\sigma_\mathcal{N} =
2.9$~deg$^{-2}$.  Therefore, the variation in the number of high
redshift galaxy clusters shows substantial cosmic
variance over fields as large as $\sim$10 square degrees.

We use the  angular correlation function, $w(\theta)$, to quantify the
clustering observed in the galaxy cluster sample, where $w(\theta)$ is
the probability of finding a companion object in a solid angle $d\Omega$ at
an angular separation $\theta$ in excess of a random distribution.
For a distribution of sources with surface density, $\mathcal{N}$,
the angular correlation function is defined as \citep[\S~45]{pee80}
\begin{equation}
dP = \mathcal{N}\,[1 + w(\theta)]\,d\Omega.
\end{equation}

The angular correlation function is calculated by comparing the total
number of source pairs at a separation $\theta$ to the expected number
of pairs at a separation $\theta$ from a random, uniformly distributed
sample.  Here we use the estimator
proposed by \citet[hereafter LS]{lan93}, which minimizes the variance and biases
associated with other estimators.   The LS estimate for the angular correlation function is
\begin{equation}\label{equation:wls}
w_\mathrm{LS}(\theta) = \frac{ DD(\theta) -2DR(\theta) + RR(\theta)} {RR(\theta)},
\end{equation}
where $DD(\theta)$ is the observed number of unique data--data pairs
with angular separation $\theta - \Delta\theta/2 < \theta < \theta +
\Delta\theta/2$, $DR(\theta)$ is the number of unique data--random
pairs in the same interval, and $RR(\theta)$ is the number of unique
random--random pairs in the same interval.  

As discussed by \citet{roc99}, because the
survey fields have a finite size the expectation value for the LS
estimator of the angular correlation function is biased to lower
amplitudes than the true angular correlation function.  It is
therefore customary to correct $w_\mathrm{LS}(\theta)$ by adding a
constant, $\mathcal{I} = w(\theta) - w_\mathrm{LS}(\theta)$.
Following \citet{qua07}, this constant is equal to the fractional
variance of the source counts,
\begin{equation}\label{equation:I}
\mathcal{I} = \frac{1}{\langle N \rangle} + \sigma^2,
\end{equation}
where $1/\langle N \rangle$ represents the Poisson error from the
finite source counts, and $\sigma^2$ is the variance arising from
object clustering in the mean density field,
\begin{equation}\label{equation:sigma}
\sigma^2 \equiv \frac{1}{\Omega^2} \iint d\Omega_1 d\Omega_2 w(\theta_{12}).
 \end{equation} 

Equation~\ref{equation:sigma} can solved numerically
if $w(\theta)$ is known \citep[\eg,][]{roc99}.   Here, we assume the
angular correlation function follows the conventional power-law,
\begin{equation}\label{equation:wtheta}
w(\theta) = A_w\theta^{-\beta}, 
\end{equation}
for which Equation~\ref{equation:sigma} becomes
\begin{equation}\label{equation:sigmab}
\sigma^2 = \frac{\sum_i  A_w\theta_i^{-\beta} RR(\theta_i)}{\sum_i RR(\theta_i)}.
\end{equation}
We solve for $A_w$ and $\beta$ using an iterative technique with
equations \ref{equation:wls}, \ref{equation:I}, \ref{equation:wtheta},
and \ref{equation:sigmab}.  In the application to the high redshift
clusters, we find that the values for $A_w$ and $\beta$ converge after
a few iterations, and the solution appears stable in either the case
where we fit for $A_w$ and $\beta$, or fit for only $A_w$ holding
$\beta$ fixed (see below). 

We estimate the uncertainty on $w(\theta)$ using two approaches.  In
the first approach, we assume the weak clustering approximation.  In
this case, the LS estimator has an uncertainty derived assuming
Poisson variance for the data--data unique pairs, $DD$,
\begin{equation}\label{equation:error}
\delta w_\mathrm{LS}(\theta) = \frac{1 + w(\theta)}
{\sqrt{DD(\theta)}}.
\end{equation}
The second approach uses the fact that we can derive the clustering in
the six SWIRE fields independently.  We then take the standard
deviation of the clustering over all fields as an estimate of the
error.    Although in practice we find that the latter uncertainty
dominates the error on the angular cluster correlation measurement, we
estimate the total uncertainty by summing these two error terms in
quadrature.

\begin{figure}[t]  
\epsscale{1.2}
\plotone{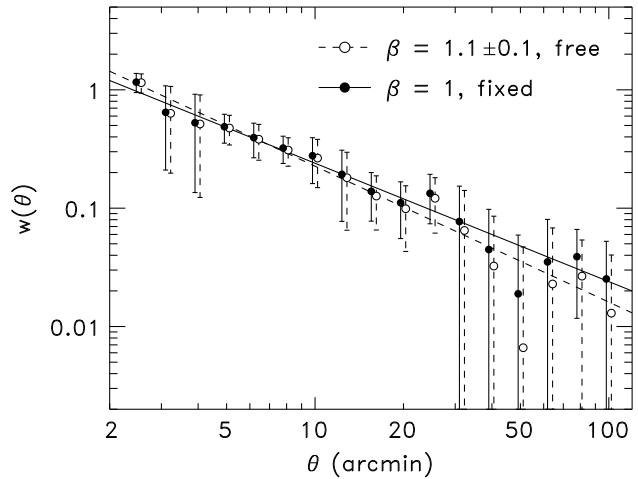}
\epsscale{1.0}
\caption{The angular correlation function for the $z \gsim 1$
clusters measured from the six SWIRE fields.  The data
points show the measured angular correlation function derived assuming
that $w(\theta) = A_w \theta^{-\beta}$, with $\beta$ as a free
parameter (filled circles and solid lines) and fixed with $\beta=1.0$
(open circles and dashed lines).  The data points are shifted slightly
along the abscissa for clarity.  The lines show the fit to
these data points over $2\arcmin < \theta < 100\arcmin$.  The difference between the data
points results from the different values for the integration constant,
$\mathcal{I}$, which depend on the values of $A_w$ and $\beta$.  \label{fig:wtheta}}
\end{figure}

We calculated the angular correlation function for the high redshift
galaxy clusters in the SWIRE fields using the above equations.  For these
calculations, we take the astrometric center of each high--redshift
cluster as the mean of the astrometric locations of all galaxies
assigned to that cluster.   We measure the angular correlation
function first for the six fields independently, and then for all six
fields combined.   Figure~\ref{fig:wtheta} shows the measured
two--point angular correlation function for the high--redshift galaxy
cluster candidates combined from all six SWIRE fields.    The error
bars on the measured amplitudes of the angular correction function
correspond to the quadrature sum of the error derived from
equation~\ref{equation:error} and the standard deviation derived from
the comparison of the six independent fields.    However, the standard
deviation between the six fields dominates the error budget.

To calculate the constant, $\mathcal{I}$, we fit the function in
equation~\ref{equation:wtheta}, with $\theta$ in units of arcminutes,
to the data over the interval $2\arcmin < \theta < 100\arcmin$, solving
first for $A_w$ and $\beta$ simultaneously, and then for only $A_w$
holding $\beta = 1.0$ fixed.  The choice of $\beta = 1.0$ follows from
measurements of the spatial correlation function of low--redshift
galaxy clusters, which generally find a slope $\beta = \gamma - 1 \approx
1.0$, where $\xi(r) = (r/r_0)^{-\gamma}$ \citep{bah03}.   The derived
value for the constant, $\mathcal{I}$, depends on the values of $A_w$
and $\beta$.   Fitting for $A_w$ and $\beta$, we obtain $\mathcal{I} =
0.04$, $A_w = 3.1\pm 0.5$, and $\beta = 1.1 \pm 0.1$.  For the case
where we hold $\beta$ fixed at 1.0, we obtain $\mathcal{I} = 0.05$ and
$A_w = 2.4\pm 0.2$.   In all cases we derive the uncertainties on
these parameters using a jackknife method
\citep[\S~6.6]{wal03}, which takes into account correlations between
the  $w(\theta)$ datapoints. Thus,  we find consistent parameters for
the power-law model in either the case where we fit for $\beta$ or
hold it fixed.

\section{Discussion}\label{section:discussion}

The angular correlation function for the high--redshift clusters is
consistent with a power-law fit over the interval $2\arcmin <
\theta < 100\arcmin$ with a power-law slope $\beta = 1$.    This slope
corresponds to a slope for the deprojected spatial correlation function,
represented by $\xi(r) = (r/r_0)^{-\gamma}$, with $\gamma = \beta + 1
= 2$.   Such a  power-law slope is representative of the correlation
function of low-redshift samples of galaxy clusters \citep[and
references therein]{bah03}.

In this section we argue that high--redshift galaxy cluster candidates
defined here correspond to the locations of mass overdensities in the
large--scale structure.    The selection of these objects using $\mone
- \mtwo > -0.1$~mag from a flux--limited catalog corresponds to a
specific range of redshift.  In this section we calculate the redshift
selection function for the high--redshift clusters.  We use this to
study the spatial clustering and space density of the high redshift
clusters, comparing them against models for the evolution of dark
matter halos.

\subsection{Redshift Selection Function}\label{section:rsf}

To derive the redshift selection function, $dN/dz$, we make the
assumption that the galaxies with $0.2 < z < 0.5$ make a negligible
contribution to the galaxy cluster selection for the following
reasons.  Firstly, it is unlikely that a galaxy cluster would exist at
$0.2 < z < 0.5$ with the requisite large number ($\geq$26 objects)
of galaxies, all with substantial emission from warm dust to satisfy
the IRAC color selection.   Secondly, the angular diameter distance,
$d_A$ at $z=0.35$ is roughly one--half that at $z=1.5$ (for
$\Omega=0.3$ and $\Lambda=0.7$).  And, by our definition (see
\S~\ref{section:selection}), the probability of identifying a $3\sigma$
overdensity in the surface density of red IRAC objects goes as
$d_A^2$.  Thus a cluster candidate composed of red IRAC objects at
$0.2 < z < 0.5$ would require $\geq$26 (dust--enshrouded,
star-forming) objects within $r\lsim
160-250$~$h^{-1}$~kpc, i.e., a physical area
$\sim$25\% relative to that for $z\gsim 1$ objects.    
Such systems should be extremely rare.   

Nevertheless, any contamination of low--redshift galaxy clusters in
the high--redshift galaxy cluster sample will suppress the intrinsic
clustering strength of the high--redshift objects.  As a further test,
we confirmed that the $\mone - \mtwo > -0.1$~mag selection excludes
the relatively low--redshift ($z\lsim 1$) X--ray selected clusters
from regions of the  XMM--LSS and ELAIS--S  SWIRE fields with
\textit{XMM} coverage \citep{pie06,puc06}.   We select none of the 20
clusters (all with $z < 1.1$) in these samples using our proposed IRAC
color selection.   Importantly, these samples include 11 clusters with
$0.2 < z < 0.5$ (where we expect some contamination of galaxies with
$\mone - \mtwo > -0.1$~mag, see \S~\ref{section:colorselection}), none
of which are identified by our cluster definition.  Therefore, we
expect a negligible contribution to the correlation function from
these low--redshift clusters.

Interesting, none of the four \textit{XMM} clusters with $z > 0.9$
(all with $z < 1.1$) would be selected by the method here.  However,
the selection method here does identify the projected high--redshift
cluster candidates at $z=1.40-1.48$ identified by \citet{vanbreukelen07}. As a
further test of the redshift distribution of the IRAC--selected
cluster candidates, we looked at the $\mone-\mtwo$ colors of the
spectroscopically confirmed cluster--member galaxies  in the IRAC
shallow cluster survey (ISCS) with $z > 1.0$ (Eisenhardt et al.\ 2007;
M.~Brodwin, private communication).   Few (only 18\%) of the
confirmed cluster--member galaxies with $z < 1.3$ have $\mone -
\mtwo > -0.1$~mag, and these clusters would likely be missed with the
selection proposed here.   However, the vast majority ($\simeq$90\%)
of the confirmed ISCS galaxies with $z > 1.3$ have $\mone - \mtwo >
-0.1$~mag.   In hindsight, the reason for this is that at $1.0 < z <
1.3$ the $\mone - \mtwo > -0.1$~mag selection misses those galaxies
with passively evolving colors, which are typical of early--type
galaxies.   As discussed in
\citet{eis07}, nearly all the confirmed cluster galaxies in their
high--redshift sample have colors consistent with older passively
evolving stellar populations.   The $\mone -
\mtwo > -0.1$~mag criterion selects galaxies with the colors of these  types of
stellar populations for $z > 1.3$ (see
figure~\ref{fig:zvcolor}). Therefore, we suspect that our color
selection is approximately complete for all galaxy clusters at $z >
1.3$.

We estimate the broad redshift selection function for the
high--redshift galaxy clusters using the observed distribution for the
galaxies selected with $\mone-\mtwo > -0.1$~mag.    We model the upper
and lower end of the redshift selection function for the
high--redshift galaxy clusters separately.  If the majority of
galaxies in clusters at $1.1 < z < 1.3$ have passively evolving
stellar populations (as for the ISCS, see above; Eisenhardt et al.\
2007), they will not be identified by our IRAC selection.
Nevertheless, we conservatively assume that the redshift selection
function for $1.1 \lsim z \lsim 1.3$ follows the distribution in
figure~\ref{fig:zdist}, and we take the spectroscopic redshift
selection function for galaxies with $R > 22.5$~mag and $\mone-\mtwo >
-0.1$~mag.  We furthermore correct the observed distribution using the
estimated spectroscopic completeness and redshift identification
completeness in the DEEP2 survey \citep{wil06}.  Because the galaxies
in all known clusters with $z < 1.1$ have colors consistent with
passively evolving populations
\citep[see][and references therein;
\S~1]{eis07}, we make the further assumption that there are no
clusters at $z < 1.1$ in our IRAC--selected sample.

To model the upper--end of the redshift selection function, we convert
the SWIRE IRAC \mone\ flux limit into the relative number of galaxies
expected per unit redshift (where again, we assume the redshift
selection function for the galaxy clusters is equal approximately to
that of the galaxies themselves).   \citet{fon06} measured the
evolution in the high--redshift galaxy mass function, parameterized as,
\begin{equation}\label{equation:massfunction}
\phi(\mu,z)\,  d\mu = 2.3\, \phi^\ast(z)\,  (
10^{\mu-\mu^\ast[z]} )^{(1+\alpha^\ast[z])} \exp(
-10^{\mu-\mu^\ast[z]} ) \, d\mu
\end{equation}
where $\mu\equiv \log(\mathcal{M})$ is the base--10 logarithm of the
stellar mass, and where $\phi^\ast (z) \equiv \phi_0^\ast(1+z)^{\phi_1^\ast}$,
$\alpha^\ast (z) \equiv \alpha_0^\ast + \alpha_1^\ast (z)$, and
$\mu^\ast (z) \equiv \mu_0^\ast + \mu_1^\ast z +
\mu_2^\ast z^2$.    Fontana et al.\ derive best--fit
parameters to their data with $\mu_0^\ast = 11.16$,
$\mu_1^\ast = 0.17$, $\mu_2^\ast = -0.07$,
$\alpha_0^\ast = -1.18$, $\alpha_1^\ast = - 0.082$, $\phi_0^\ast =
0.0035$, and $\phi_1^\ast = -2.20$.     We calculate the number density
of galaxies above some stellar  mass limit and redshift integrating
equation~\ref{equation:massfunction}, 
\begin{equation}\label{equation:nmass}
n(>\mathcal{M},z) = \int_{\log \mathcal{M}}^{\infinity} \phi(\mu, z)\, d\mu.
\end{equation}

We convert equation~\ref{equation:nmass} into the number of galaxies
above the IRAC flux limit for SWIRE as a function of redshift using an
estimate for the galaxy mass--to--light ratio, $\mathcal{M}/L_\nu(z)$,
where $L_\nu(z)$ is the luminosity density at 3.6/(1+$z$)~\micron.
Here, we take $f_\mathrm{lim}(3.6\micron) = 7$~\uJy\ as the flux
limit.   The observed mass--to--light ratio depends on the
relative number of early-- and late--type stellar populations
constituting each galaxy
\citep[\eg,][]{bru03} and dust extinction.    However, \citet{rud06}
showed that the global galaxy population at $0.7 < z < 2.5$ has
rest--frame $U-B$ and $B-V$ colors consistent with a simple
exponentially declining star-formation rate with a characteristic
timescale of 6~Gyr, formed at $z_\mathrm{form} = 4$ with dust
extinction of $A_V = 0.6$~mag.   We therefore use a stellar population
model \citep[from][]{bru03} with these parameters to compute
$\mathcal{M}/L_\nu(z)$.  The limiting mass as a function of redshift
is then $\mathcal{M}_\mathrm{lim} = \mathcal{M}/L_\nu(z) \times
f_\mathrm{lim}(3.6\micron) \times 4\pi D_L^2(z) (1+z)^{-1}$, where
$D_L(z)$ is the luminosity distance.  We insert this expression  into
equation~\ref{equation:nmass}.   The redshift distribution function is
then the derivative of equation~\ref{equation:nmass} with respect to
redshift, $dN/dz = dn(>\mathcal{M}_\mathrm{lim},z)/dz$.\footnote{If
cluster galaxies have higher mass-to-light ratios than the global
galaxy at all high redshifts, then the redshift selection function
would underpredict the number of high--redshift clusters.  However,
the difference in $\mathcal{M}/L_\nu(z)$ at 3.6/$(1+z)$~\micron\
between the global galaxy population used here and galaxies with a
maximally high mass-to-light ratio (i.e., corresponding to a stars
formed in a instantaneous burst at $z_\mathrm{form}=\infinity$) is
less than a factor of 1.5 for redshifts of interest, $1 < z <
2.5$.   Therefore, even if this scenario occurs it will have only a
small effect on the redshift selection function.}

\begin{figure}[t]  
\epsscale{1.2}
\plotone{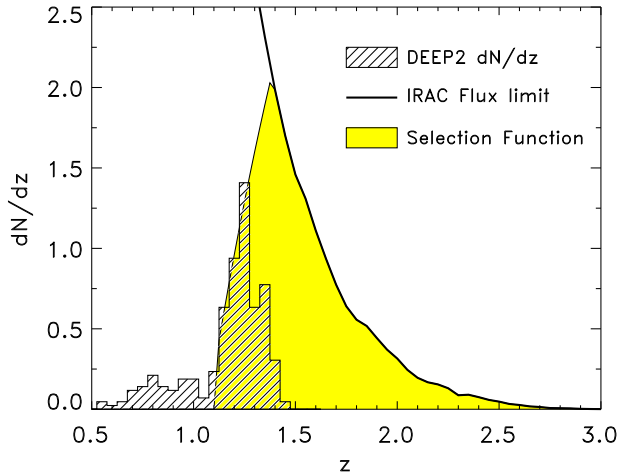}
\epsscale{1.0}
\caption{The redshift selection function for high redshift galaxy
clusters selected from the SWIRE data with $\mone - \mtwo > -0.1$~mag.
The hashed histogram shows the spectroscopic redshift distribution of
DEEP2 galaxies with $\mone - \mtwo > -0.1$~mag, $R > 22.5$~mag, and
IRAC flux densities above the SWIRE limit, which represents the
lower--end of the redshift selection function.  The thick solid line shows
the expected redshift distribution of IRAC sources to the SWIRE
flux limit with assumptions on the evolution of the mass function and
color distribution of galaxies (see text), and this represents the
upper--end of the redshift selection function.   The intersection of the
lower-- and upper--end of the redshift selection functions provides
the total redshift selection function, $dN/dz$, which is indicated by
the yellow--shaded region.
\label{fig:dndz}}
\end{figure}

The intersection of the lower-- and upper--end of the redshift
selection functions yields the total redshift selection function.
Figure~\ref{fig:dndz} shows the  total redshift selection function and
the individual components.  We normalize the redshift selection
function with the convention, $\int dN/dz\;\; dz = 1$.   The mean and
variance of the redshift distribution function are the first and
second moments of this distribution, which give $\langle z \rangle =
1.54$ and $\sigma(z) = \sqrt{\sigma^2} = 0.28$.

We use the redshift selection function to estimate the spatial number
density of the galaxy clusters.   The number density is equal to
\begin{equation}\label{equation:ndens}
n = \frac{\mathcal{N}}{ \int (1+z)^2\, p(z)\, d_A^2\, E(z)^{-1}\, dz}, 
\end{equation}
where $\mathcal{N}$ is the observed surface density, and we define the function \citep[pgs.~100, 312, 332]{pee93}
\begin{equation}\label{equation:efunc}
E(z) = \left( \frac{c}{H_0} \right)^{-1} [ \Omega_m(1+z)^3 + \Omega_R
(1+z)^2 + \Omega_\Lambda]^{1/2}, 
\end{equation}
where $\Omega_R = 1 - \Omega_m - \Omega_\Lambda \equiv 0$ here.
The denominator of equation~\ref{equation:ndens} is the effective
volume per unit solid angle.  The effective volume differs from the
comoving volume by $p(z)$, which is the probability that a cluster
with redshift $z$ will be selected using the method here, thus $0 \leq
p(z) \leq 1$ for all $z$.   The quantity $p(z)$ compensates for
various selection biases and incompleteness.   We generally assume
here that $p(z) \propto dN/dz$ with a normalization such that $p(z) =
1.0$ for $1.3 < z < 1.5$.     However, the completeness is poorly
known given the relatively complicated selection function for the high--redshift
clusters (see above).  Therefore,
in the discussion that follows we also consider other distributions for
$p(z)$ that should span the possible plausible range.  This provides
limits on the effective volume for the high--redshift clusters until
improvements in the selection function become available (either from
spectroscopic or accurate photometric redshifts).

Applying equation~\ref{equation:ndens} to the redshift
selection function with the numbers in table~\ref{table}, the spatial
number density of the high redshift galaxy clusters is $n = 1.2\pm 0.1
\times 10^{-5}\, h^3$~Mpc$^{-3}$, where the uncertainty is the standard
deviation on the spatial densities derived separately for the six
SWIRE fields.  However, the number density  dependent on the assumed
redshift selection function, which we discuss further in \S~5.3.

\subsection{Spatial Clustering of High Redshift Clusters}

The angular correlation function, $w(\theta)$,  corresponds to the 3--dimensional
spatial correlation function, $\xi(r)$, projected on sky.  They are
related through the Limber projection using a known redshift selection
function, $dN/dz$ \citep[\S~50, 52]{efs91,pee80}.   Allowing the
evolution of the spatial correlation function to follow, $\xi(r,z) =
\xi(r,0) \times f(z)$, where $\xi(r,0) = (r/r_0)^{-\gamma}$ as
above,  and conventionally $f(z) = (1+z)^{-(3-\gamma+\epsilon)}$, then the
relation between the amplitude of the angular correlation function and
the spatial correlation function is \citep{efs91}
\begin{equation}\label{equation:limber}
A_w = H_\gamma\, r^\gamma_0\, \int f(z)\, d_C^{1-\gamma}(z)\, \left( \frac{dN}{dz}
\right)^2 E(z) dz \,\, \left( \int \frac{dN}{dz} dz \right)^{-2},
\end{equation}
where $d_C$ is the comoving distance, $dN/dz$ is the redshift
selection function, $E(z)$ is defined in equation~\ref{equation:efunc},
and $H_\gamma$ is a numerical factor given by \citep[\S~52]{pee80}
\begin{equation}
H_\gamma = \sqrt{\pi} \frac{\Gamma[(\gamma-1)/2]}{\Gamma(\gamma/2)}.
\end{equation}

\begin{figure}[b]  
\epsscale{1.2}
\plotone{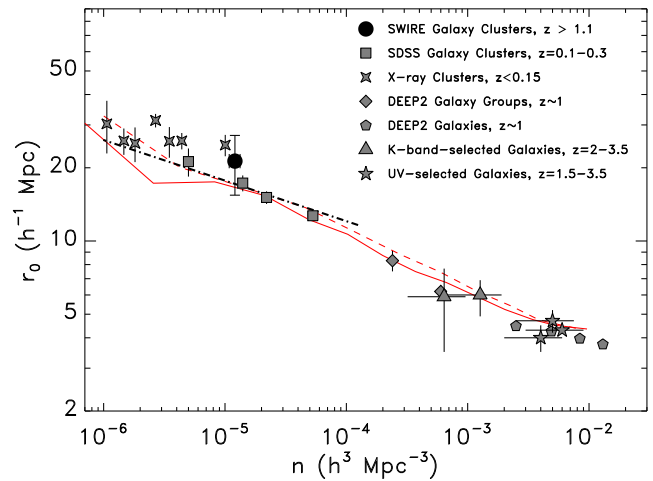}
\epsscale{1.0}
\caption{ The comoving number density, $n$, versus the spatial correlation function
scale length, $r_0$, for various objects at different redshifts.  The
large, filled circle shows the value derived here for the
IRAC--selected high--redshift cluster candidates, assuming the redshift distribution
in figure~\ref{fig:dndz}, and includes an additional systematic error (summed in
quadrature) of $\delta(r_0) = 5$~$h^{-1}$ Mpc for uncertainties in the
redshift selection function.   The filled squares show optically-selected
galaxy clusters from SDSS with $N_\mathrm{gal} > 10$, 13, 15, and 20
from \citet{bah03}.   The four--point stars show X-ray--selected
galaxy clusters with $L_X \gsim 10^{43}$~erg~s$^{-1}$
\citep{aba98,lee99,col00,bah03}.  Filled pentagons
show the $z\sim 1$ galaxy sample from DEEP2 \citep{coil06a}, and
filled diamonds show the $z\sim 1$ galaxy--group sample from DEEP2
\citep{coil06b}.   The filled triangles show $K$--band
selected galaxies at $z\sim 2$ from \citet{qua07}.  The filled
pentagrams show UV--selected ``U--dropouts'' at $1.5 < z < 2.0$, $2.0
< z < 2.5$, and $2.5 < z < 3.5$ from \citet{ade05b}.  The lines show
various models.  The dot--dashed line shows the empirical
relation $r_0 \sim 2.6n^{-3/2}$ derived for a $\Lambda$CDM model used by \citet{bah03}.   The red
lines show predictions from the Millennium model simulations for
$z=0.2$ (solid line) and $z=1.5$ (dashed line). \label{fig:r0dens}}
\end{figure}

Following the arguments of \citet{gia98}, for the large spatial scales
considered here the effective variation in the correlation length,
$r_0$, should be small.  Therefore, we take $\epsilon=\gamma-3$,
corresponding to constant clustering in comoving units over the
redshift range considered here.  In this case, $r_0(z) = r_0$ is the
correlation length at the epoch of the observation. 

We solve equation~\ref{equation:limber} for the spatial correlation
scale length, $r_0$, using the redshift
selection function derived in \S~\ref{section:rsf} (see
figure~\ref{fig:dndz}), and $A_w$ and $\beta$ derived for the angular
correlation function in \S~\ref{section:wtheta}.   For the case where
$\beta=1.0$ ($A_w = 2.4\pm 0.2$), we derive $r_0 = 26.9 \pm 5.6$
$h^{-1}$~Mpc.  For the case where $\beta$ varies ($A_w = 3.1\pm 0.5$
and $\beta=1.1\pm0.1$), we derive $r_0 = 22.4 \pm 3.6$ $h^{-1}$~Mpc.
Because these two cases give consistent answers, we quote here the
value for the latter, which includes the larger error and is thus more
conservative.    

Figure~\ref{fig:r0dens} illustrates the relation of $r_0$ to the
spatial density, $n$, for the high--redshift galaxy clusters discussed
here to other samples in the literature.     The spatial correlation
scale length for the high--redshift galaxy clusters is comparable to
that derived for optically--selected rich clusters at relatively low
redshift \citep[$0.1 < z < 0.3$,][]{bah03}.   This implies that the high--redshift galaxy
clusters selected by IRAC are progenitors of low--redshift galaxy
clusters.   The spatial correlation function scale length is also
consistent with those derived for luminous X-ray clusters ($L_X \gsim
10^{43}$~erg~s$^{-1}$), which range from $r_0 \sim 25-30$~$h^{-1}$ Mpc
\citep[with values taken from Bahcall et al.\
2003]{aba98,lee99,col00}.   Therefore, some of the high--redshift
galaxy clusters seem destined to become luminous X--ray clusters.

The largest uncertainty in the derived spatial correlation scale
length stems from the assumed shape of the redshift selection function,
$dN/dz$.  In particular, our redshift selection function is fairly
broad and assumes the clusters are smoothly distributed over $1.1 < z
\lsim 3.0$.   If the clusters show prominent, discrete voids and
spikes in this distribution, this will lower the spatial correlation
scale length \citep{ade05a}.   However, because we average the
correlation functions over the six separate SWIRE fields, we expect
this effect is less severe.  Larger uncertainties likely result from
the cluster galaxy colors and from the fact that the redshift
distribution of bona fide high--redshift galaxy clusters may not be
equal to that of the general galaxy population.   For
example, one extreme (yet very possible) scenario is that galaxies in
all clusters with $1.1 < z < 1.5$ have colors consistent with
passively evolving stellar populations.   In this case, the IRAC
$\mone - \mtwo > -0.1$~mag selection would miss all clusters with $z <
1.3$, like those in the ISCS discussed above \citep[see
\S~\ref{section:rsf}; ][]{eis07}.   Furthermore, if the redshift
distribution of galaxy clusters evolves more strongly at high
redshifts than the general galaxy population, then the redshift
selection function in figure~\ref{fig:dndz} would over predict the
number of clusters with $z \gsim 1.5$.   

All the effects discussed above would narrow the redshift selection
function compared to that derived in \S~\ref{section:rsf}, which would
\textit{lower} the derived spatial correlation function scale length.    As a
fiducial example, inserting a redshift selection function described by
a Gaussian, $dN/dz \propto \exp( -(z - \bar{z})^2 / 2\sigma_z^2)$ with
$\bar{z} = 1.5$ and $\sigma_z = 0.1$, would reduce the spatial
correlation function scale length to $r_0 \approx 18$~$h^{-1}$~Mpc
(consistent with the $r_0$ derived for $\langle z \rangle = 1$
clusters in the ISCS, see Brodwin et al.\ 2007).    Therefore, we incorporate the
uncertainty on the redshift selection function by adding (in
quadrature) an additional error $\delta(r_0) = 5$~$h^{-1}$~Mpc to the
correlation length in figure~\ref{fig:r0dens}.  However, this error is
systematic, and its general effect is to lower the derived value of
$r_0$.   Such a redshift
selection function would also increase the number density derived in
equation~\ref{equation:ndens} to $\approx 2 \times 10^{-5}$~$h^{3}$~Mpc$^{-3}$.
To remove this source of systematic uncertainty requires
deriving a more accurate redshift selection function, either through
spectroscopy or with good photometric redshifts.  The data required to
estimate a more accurate redshift selection function do not yet exist
over the full SWIRE fields.

Current large--scale cosmological simulations using an $\Lambda$CDM
model reproduce the comoving clustering and space density of the
galaxy clusters, groups, and galaxies.   Figure~\ref{fig:r0dens} shows
a fit to the $\Lambda$CDM model invoked by \citet{bah03}, $r_0 \sim
2.6n^{-3/2}$, over the range $20 < n^{-1/3} < 90\, h^{-1}$~Mpc.   This
model intersects the optically-selected clusters, although it corresponds
to values or $r_0$ lower than those measured for the less-luminous
X-ray clusters.   Similarly, recent cosmological models from the
Millennium simulation \citep{spr05} reproduce the data on scales from
galaxy clusters to the galaxies themselves.  The two curves in the
figure show the Millennium--simulation predictions for $z=0.2$ and
1.5.  The model predictions show the relationship between the
clustering strength and space density of dark--matter halos.
Therefore, the measured clustering of galaxy clusters, galaxy groups, and
galaxies implies that these objects trace the underlying dark matter
halos over a large range of mass scale and redshift.

\begin{figure}[t]
\epsscale{1.2}
\plotone{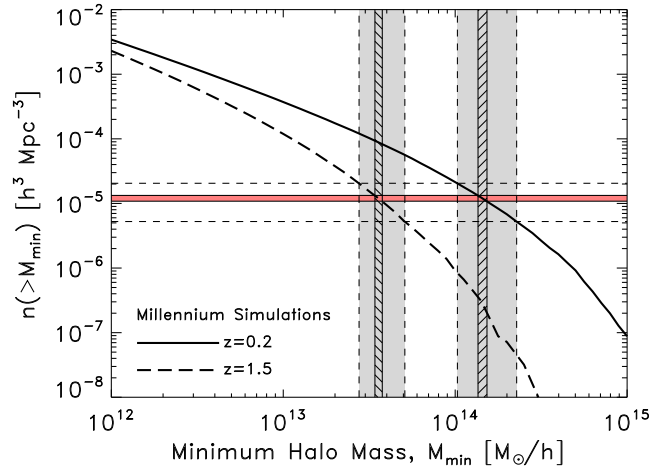}
\epsscale{1.0}
\caption{Number density of dark matter halos as a function of halo
mass.  The curves show the expected number density of halos greater
than mass, $\mathcal{M}$, for $z=0.2$ (solid lines) and $z=1.5$ (dashed lines)
from the Millennium model simulations \citep{spr05}.  The shaded
horizontal band shows the measured number density of galaxy clusters
at $z\gsim 1$ from SWIRE assuming the fiducial redshift selection
function derived in \S~5.1.   The horizontal short-dashed
lines show the bound on the number density for limiting cases in the
assumed redshift selection function.
The vertical hashed regions show masses to which the measured number density corresponds
at each redshift, and the shaded region extends these for the limiting
number densities.  While the measured number density corresponds to a
halo mass of $\approx 3-5\times 10^{13}\; h^{-1}$~$\mathcal{M}_\odot$ at $z=1.5$, these
will evolve to have halo masses of $\approx 1-3\times
10^{14}\; h^{-1}$~$\mathcal{M}_\odot$ at $z=0.2$. \label{fig:nmass}}
\end{figure}

\subsection{Evolution of Galaxy Clusters}

The cosmological simulations match the behavior of the number density
and spatial correlation function scale length of the high--redshift
galaxy clusters.  Thus, we may relate the derived number density for
the high--redshift galaxy clusters to estimate the corresponding dark
matter halo mass in the simulations.   Figure~\ref{fig:nmass} shows
the expected number distribution of dark--matter halos above a given
mass threshold at $z=1.5$ and 0.2 for the Millennium simulation
\citep{spr05}.   At $z=1.5$, near the mean redshift of our redshift
distribution function (\S~\ref{section:rsf}), the comoving number
density of the IRAC--selected high--redshift galaxy clusters
corresponds to dark matter halos larger than $3.5\times
10^{13}$~$h^{-1}$~\msol.   

As discussed in \S~5.2, the number
density is uncertain due to the unknown redshift selection function.
In the discussion here, we allow for two limiting redshift selection
functions to constrain the range for the comoving number density, and
thus the range of dark matter halo masses.  The first limiting number
density corresponds to the case where the redshift selection function
is represented by a Gaussian with $\langle z \rangle$=1.5 and
$\sigma_z = 0.1$.  In this case the number density would increase to
$2 \times 10^{-5}$~$h^3$~Mpc$^{-3}$.  Alternatively, we consider the
case where the redshift selection function is constant for $1.1 < z <
2.5$ and zero elsewhere.  In this case, the number density would
decrease to $3\times 10^{-6}$~$h^3$~Mpc$^{-3}$.  These cases are
arguably limiting cases, and thus they bound the range of possible
comoving number densities.  Figure~\ref{fig:nmass} shows the number
densities for these cases, and these would correspond to dark matter
haloes at $z=1.5$ larger than $\approx$$3-5\times 10^{13}$~$h^{-1}$~\msol.  

Due to hierarchical nature of CDM, these
objects continue to accrete matter with decreasing redshift.
Figure~\ref{fig:nmass} shows that the range of comoving number
densities for the
IRAC--selected galaxy clusters at $z\gsim 1$ correspond to dark matter
halos larger than $1-3\times10^{14}$~$h^{-1}$~\msol\ at $z=0.2$, which is
comparable to the dark--matter halo masses of the SDSS
optically-selected clusters \citep{bah03} and X-ray clusters
\citep{aba98,col00}.

Therefore, the high--redshift galaxy clusters identified by the $\mone
- \mtwo > -0.1$~mag selection correspond to rich present--day
clusters.  This is perhaps unsurprising given that we defined the
high--redshift clusters to have $\geq$26--28 companions within angular
sizes of 1.4~arcminutes.   Presumably the number density (and spatial
clustering correlation length) depends on this definition.  Requiring
a larger number of galaxies within this aperture would identify rarer
objects with a lower number density, perhaps identifying the denizens
of larger--mass dark matter halos.  Similarly, requiring a smaller
number of galaxies presumably corresponds to lower--mass halos,
although using a smaller number of galaxies will also suffer more from
contamination from the average surface density for objects with $\mone
- \mtwo > -0.1$~mag unassociated with galaxy clusters.

It is worth noting that the high--redshift galaxy clusters defined
here are relatively unbiased by the (rest--frame) optical colors of the
galaxies.  They have no dependence on any color other than $\mone -
\mtwo$.  Therefore, measuring the distribution of rest--frame optical colors in
the galaxies of these clusters will allow the study of when and how
these objects formed their stellar populations.  However, to derive
accurate rest--frame colors requires more accurate redshift
estimates than what is possible with only the $\mone-\mtwo$ color and
the redshift selection function in \S~\ref{section:rsf}, and good
redshifts (photometric or spectroscopic) are needed for further study
of the colors of the galaxies in these clusters.  Unfortunately, the
current optical imaging with the public SWIRE data release covers only
$\approx$30\% of the area of the IRAC survey (and not all of these
data are useful for photometric redshifts).   More ancillary data will
be required to study the evolution of the galaxies in the
high--redshift IRAC--selected clusters, and to compare them to the
galaxies in lower redshift galaxy clusters.

\section{Summary}

We discuss the angular clustering of galaxy clusters at $z > 1$
selected over 50~deg$^2$ from SWIRE.  We select high--redshift
galaxies ($z \gsim 1$) using a simple color selection, $\mone - \mtwo
> -0.1$~mag.   The small number of contaminants may be rejected using
an additional apparent magnitude limit $R > 22.5$~mag, which
efficiently removes galaxies with $0.2 < z < 0.5$.

From the galaxies with $\mone - \mtwo > -0.1$~mag, we identify galaxy
cluster candidates as objects with $\geq$26--28 companions within $r =
1\farcm4$ radii.   Using datasets with high--quality spectroscopic
redshifts, we show that the majority ($>$80\%) of all galaxies
satisfying $(\mone -\mtwo)_\mathrm{AB} > -0.1$~mag have $z > 1.0$.
Furthermore, more than 50\% (90\%) of the galaxies with spectroscopic
redshifts $z > 1.1$ (1.3) satisfy this IRAC color selection.  

These candidate galaxy clusters show strong angular clustering.  From
the data, we derived an angular correlation function represented by
$w(\theta) = (3.1\pm0.5) (\theta/1\arcmin)^{-1.1\pm0.1}$  over scales
of 2--100~arcmin.   The slope of the angular correlation function,
$\beta$, corresponds to a slope of the spatial correlation function
$\gamma =\beta + 1 \simeq 2.0$, consistent with the slope for
low--redshift galaxy clusters.

Assuming the redshift distribution of these galaxy clusters follows
our fiducial model, these galaxy clusters have a spatial--clustering
scale length $r_0 = 22.4\pm 3.6 h^{-1}$~Mpc, and a number density $1.2\pm 0.1
\times 10^{-5} h^{3}$~Mpc$^{-3}$.    The correlation scale length and
number density of these objects are comparable to those of rich
optically--selected and X-ray--selected galaxy clusters at low
redshift.   However, the largest uncertainty on the
spatial--clustering scale length stems from uncertainties in the
redshift selection function.     The $\mone - \mtwo > -0.1$~mag
selection would miss the galaxies in clusters at $1.1 < z < 1.3$, if
these clusters are dominated by galaxies  with colors consistent with
passively evolving stellar populations.   Furthermore, if the redshift
distribution of galaxy clusters evolves more strongly at high
redshifts than the general galaxy population, then our redshift
selection function would over predict the number of clusters at the highest
redshifts.  These effects would narrow the redshift selection
function, which would tend to lower the spatial
correlation function scale length.  To improve the measurement on the
spatial correlation function requires a more accurate redshift selection
function, either from deep spectroscopy or well--calibrated photometric
redshifts.

Comparing the number density of these high--redshift clusters to
dark--matter halos from the $\Lambda$CDM Millennium simulations, the
high--redshift clusters correspond to dark--matter halos larger than
$3-5\times
10^{13}$ $h^{-1}$ \msol\ at $z=1.5$, including an allowance for the
possible range of number densities.   Assuming these grow  following
$\Lambda$CDM models, these clusters will reside in halos larger than $1-3\times
10^{14}$ $h^{-1}$ \msol\ at $z=0.2$, comparable to large galaxy
clusters at low redshift.   Therefore, the IRAC--selected galaxy
clusters correspond to the high--redshift progenitors of present--day
galaxy clusters.

The selection of the high--redshift galaxy clusters has no dependence
on the rest--frame optical colors of the galaxies themselves.
Therefore, future observations of the galaxies in these high--redshift
galaxy clusters allows the study of their star--formation histories
with little additional bias.

\acknowledgments

The author acknowledgments many invaluable discussions with colleagues
that led to the analysis and interpretation in this paper.   In
particular, the author would like to thank Mark Brodwin, Alison Coil,
Asantha Coorary, Romeel Dav\'e, Daniel Eisenstein, Harry Ferguson,
Mauro Giavalisco, Ivo Labb\'e, Ivelina Momcheva, Roderik Overzier,
Marcia Rieke, Gregory Rudnick, Suresh Sivanandam, Daniel Stern, Pieter
van Dokkum, Benjamin Weiner, Christopher Willmer, and Andrew Zirm.
The author also acknowledges very helpful comments and corrections
from the anonymous referee.  The author thanks the SWIRE Legacy team
for producing a high--quality dataset, and the author thanks the
Millennium collaboration for making their simulations available. This
work is based on data obtained with the Spitzer Space Telescope, which
is operated by the Jet Propulsion Laboratory (JPL), California
Institute of Technology (Caltech) under a contract with NASA.  Support
for this work was provided by NASA through the Spitzer Space Telescope
Fellowship Program, through a contract issued by JPL, Caltech under a
contract with NASA.

\end{document}